\documentclass[12pt]{article}

\begin{document}

\hspace*{-15mm}\begin{minipage}{145mm}
\title{Position eigenstates, symmetries, and the redundant hermiticity 
of free-particle Hamiltonians 
\vspace{10mm}}
\author{{\bf L. Polley}\\[5mm]Physics Department\\ 
                   Oldenburg University\\
		   26111 Oldenburg, FRG\vspace{15mm}}
\maketitle
\vspace*{12mm}
\thispagestyle{empty}
\abstract{The quantum state of a particle can be completely
specified by a position at one instant of time. This implies a lack 
of information, hence a symmetry, as to where the particle will move. 
We here study the consequences for free particles of spin 0 and spin 1/2. 
On a cubic spatial lattice a hopping equation is derived, and the continuum limit 
taken. 
Spin 0 leads to the Schr\"odinger equation, and spin 1/2 to the Weyl equation.
Both Hamiltonians are hermitian automatically, if time-reversal symmetry is assumed. 
Hopping amplitudes with a ``slight'' inhomogeneity lead to the Weyl equation in a 
metric-affine space-time.
 }
\end{minipage}\newpage

For a state vector which is a superposition of eigenvectors of an
observable, the probability for measuring an eigenvalue is $|\psi_i|^2$ if  
$\psi_i$ is the corresponding amplitude. The prevailing view 
about this rule is that it must be taken as an independent Statistical Axiom.   
Several authors, however, have recently presented arguments to show that the relation 
of amplitudes and probabilities is contained already in the non-statistical axioms of 
quantum mechanics \cite{Deutsch1,Deutsch2,DeWitt,redund}, thus raising an issue of 
potential interest relating to the difficulties in defining a satisfactory scalar 
product for state vectors in General Relativity \cite{Ashtekar}. 
Clearly, there were assumptions to be made. 
In \cite{Deutsch1,Deutsch2,DeWitt} the normalization of state 
vectors in the usual sense was anticipated; in \cite{redund} it was assumed that in an 
isolated two-state system there are no ``decaying'' modes of evolution. The present 
paper intends to weaken the latter assumption (thus generalizing the conclusion) 
from ``non-decaying'' to ``non-singular''. Instead of two-state systems, we here 
consider free particles of spin 0 and of spin 1/2. For these, the unitarity of 
time evolution can be inferred from merely
\begin{itemize}
\item the superposition principle 
\item symmetry under translation, rotation, and time-inversion 
\item the ``logic'' of position eigenstates.      
\end{itemize}
The argument is very simple for a scalar particle living on the sites
of a cubic lattice. The particle's position at the instant of preparation already 
specifies the state vector (up tp a phase), so there is an unavoidable indeterminacy 
as to where the particle will go next. On a lattice, all nearest neighbours are 
equivalent destinations of the hopping process. This amounts to a symmetry which is 
not present in classical motion. Indeed, only by forming superpositions---the unique 
quantum-mechanical concept relating to undecided alternatives---can a particle 
``move'' out of a position eigenstate. The symmetry among the nearest neighbours in 
this process allows to reconstruct the one-particle Hamiltonian without any reference
to the classical equations of motion, as elaborated already in \cite{HupfQED}. In the 
present paper, moreover, we avoid anticipation of a scalar product, invoking higher 
symmetries and general superpositions instead.  

The argument takes its most suggestive form in the Heisenberg picture.        
Let $a$ be the spacing of a cubic lattice and let
$$
     |\vec{x},t\rangle   \qquad \vec{x} = a\vec{n}, \quad \vec{n}\mbox{ integer}
$$  
be an eigenstate of the position operator, prepared at the time $t$. When this
same state vector is expressed in terms of position eigenstates referring to a
time infinitesimally different, $t'=t+{\rm d}t$, we expect to be dealing with a 
superposition of nearest-neighbour states, i.e. 
\begin{equation}    \label{hop0}
    |\vec{x},t\rangle = \alpha |\vec{x},t+{\rm d}t\rangle + 
 \sum_{{\rm nearest}\atop{\rm neighbours}} \beta |\vec{x}+a\hat{n},t+{\rm d}t\rangle  
\end{equation}
where $\alpha$ and $\beta$ are some complex numbers dependent on the time step
${\rm d}t$. For ${\rm d}t\to 0$ we expect $\alpha\to 1$ and $\beta\to 0$; hence, 
$$
    \alpha = 1 + \epsilon {\rm d}t + {\cal O}({\rm d}t^2) \qquad \qquad
    \beta  = \kappa {\rm d}t + {\cal O}({\rm d}t^2)
$$
Thus (\ref{hop0}) amounts to the differential equation
\begin{equation}    \label{hop1}
  - \frac{{\rm d}}{{\rm d}t} |\vec{x},t\rangle = \epsilon  |\vec{x},t\rangle +   
    \sum_{{\rm nearest}\atop{\rm neighbours}} \kappa |\vec{x}+a\hat{n},t\rangle          
\end{equation}
Invoking the superposition principle again, we now consider a general state vector 
$|\psi\rangle$ in the Heisenberg picture. 
It is time-independent, or rather ``timeless'' \cite{Rovelli} as 
a matter of concept, but $|\psi\rangle$ can be expressed in a basis of position 
eigenstates referring to any time $t'$. Thus   
\begin{equation}    \label{genSup} 
    |\psi\rangle = \sum_{\rm lattice\atop sites} \psi(\vec{x},t') |\vec{x},t'\rangle
\end{equation}
where $\psi(\vec{x},t')$ denote some complex coefficients. Differentiating
(\ref{genSup}) with respect to $t'$, using the linear independence of 
$|\vec{x},t'\rangle$ at any fixed $t'$, and using equation (\ref{hop1}), 
we conclude that
\begin{equation}    \label{SchrDiskr}
    \frac{{\rm d}}{{\rm d}t'} \psi(\vec{x},t') =  \epsilon \psi(\vec{x},t') + 
    \sum_{\rm nearest\atop neighbours} \kappa \psi(\vec{x}+a\hat{n},t')
\end{equation} 
In order to take the continuum limit, which physically could mean something
like $a\to a_{\rm Planck}$ rather than $a\to0$, we expand 
$$
    \psi(\vec{x}+a\hat{n},t') = \psi(\vec{x},t') 
        + a\hat{n}_i\nabla_i\psi(\vec{x},t')
  + {\textstyle\frac12} a^2 \hat{n}_i \hat{n}_j \nabla_i \nabla_j\psi(\vec{x},t')
        + {\cal O}(a^3) 
$$
In the sum over nearest neighbours in equation (\ref{SchrDiskr}) the first-derivative
terms cancel out while second-derivative terms remain to form the Laplacian of $\psi$.
Thus, changing $t'$ into $t$ again,  
$$
    \frac{\partial}{\partial t} \psi(\vec{x},t) = (\epsilon+6\kappa)\psi(\vec{x},t)
    + a^2 \kappa \Delta\psi(\vec{x},t) + {\cal O}(a^3)
$$ 
When terms of ${\cal O}(a^3)$ are omitted, the general solution of the equation is
\begin{equation}    \label{genSol}
    \psi(\vec{x},t) = e^{(\epsilon+6\kappa)t} \int \tilde{\psi}(\vec{p}) 
                e^{i\vec{p}\cdot\vec{x}} e^{-\kappa a^2 \vec{p}^2 t} {\rm d}^3p
\end{equation} 
where $\tilde{\psi}(\vec{p})$ is the Fourier transform of the wave function at
$t=0$. Assuming that quantum mechanics has a {\em time-reversal symmetry} 
we can see from (\ref{genSol}) that hopping parameter $\kappa$ must be purely 
imaginary. This is because $\Re\kappa\neq0$ would imply an evolution factor 
exponentially increasing like 
$$
     \exp\left(|\Re\kappa|a^2 \vec{p}^2 |t| \right)
$$
in either the positive or negative time direction. Thus any wave function at $t=0$ 
with a fall-off in momentum space slower than Gaussian would run into a singularity 
immediately at either $t<0$ or $t>0$. The other evolution factor in equation
(\ref{genSol}), $e^{(\epsilon+6\kappa)t}$, is independent of the one-particle state 
vector $|\psi\rangle$. It would, however, multiply to the $n$th power in a state
with $n$ particles. Hence it would lead to a singularity similar to the above for any 
superposition of $n$-particle states with a fall-off in $n$ slower than exponential.
Therefore $\epsilon$, too, must be purely imaginary. Thus $e^{(\epsilon+6\kappa)t}$ 
can be absorbed in a universal redefinition of the phase of the wave functions.

The argument does not depend on the number of spatial dimensions. It can 
be used to corroborate the ``no decay'' assumption of \cite{redund}, relating to the 
propagation of a quantum particle through a ``channel'' connecting ``cavities''. 
If the ``channel'' is modelled by a one-dimensional array of lattice sites, then
in the continuum limit the propagation is unitary automatically.   

Let us now consider a spin 1/2 particle living on the sites of a cubic spatial
lattice. A similar scenario was investigated by Bialynicki-Birula \cite{Bialyn}
in connection with cellular automata; our approach differs by letting time run 
continuously, and by not anticipating the unitarity of the time evolution. Let 
$$
  |\vec{x},s,t\rangle \qquad \vec{x}=a\vec{n} \qquad s = \pm{\textstyle\frac12}
$$
be an eigenstate of position and spin prepared at the time $t$. In terms of
state vectors prepared at $t'= t+{\rm d}t$ we expect a superposition involving   
nearest-neighbours and spin-flips. 
As in the scalar case, let nearest-neighbour directions be given by unit vectors 
$\hat{n}$ with $n=\pm1,\pm2,\pm3$, where $\widehat{-n}=-\hat{n}$. For convenience, 
let $\hat{0}=0$ denote the unshifted position. We then expect a hopping equation of the from
\begin{equation}      \label{HopSpin}
 |\vec{x},s,t\rangle = \sum_{n=-3}^3 c_{nss'} |\vec{x}+a\hat{n},s',t+{\rm d}t\rangle
\end{equation} 
where we adopt the {\em summation convention} with respect to the spin. In the limit 
${\rm d}t\to 0$ the {\sc rhs} must tend to the {\sc lhs}, so
\begin{eqnarray*}
   c_{0ss'} &=& \delta_{ss'} + H_{0ss'} \, {\rm d}t + {\cal O}({\rm d}t^2)  \\
   c_{nss'} &=& H_{nss'} \, {\rm d}t + {\cal O}({\rm d}t^2)  
                     \qquad  n=\pm1, \pm2, \pm3
\end{eqnarray*}
Thus (\ref{HopSpin}) amounts to the differential equation 
\begin{equation}      \label{HoppDiff}
  \frac{\rm d}{{\rm d}t} |\vec{x},s,t\rangle =  
  \sum_{n=-3}^3 H_{nss'} |\vec{x}+a\hat{n},s',t\rangle    
\end{equation}
Now let $|\psi\rangle$ be a general, time-independent state vector, and express it 
as a superposition
$$
 |\psi\rangle = \sum_{\vec{x}} \psi_s(\vec{x},t) |\vec{x},s,t\rangle
$$
By the requirement that $|\psi\rangle$ be constant while the choice of basis  
vectors is changing, the time-dependence of the coefficients $\psi_s(\vec{x},t)$ 
is determined:
\begin{equation}   \label{PsiDiff}
  \frac{\rm d}{{\rm d}t} \psi_s(\vec{x},t) =  
     \sum_{n=-3}^3 H^T_{nss'} \psi_{s'}(\vec{x}-a\hat{n},t)
\end{equation}
We now evaluate the symmetries characterizing a {\em free} spin 1/2 particle.
Translational invariance is already incorporated in (\ref{PsiDiff}). 
As for invariance under lattice rotations, let direction $n$ be rotated into 
direction $m$, and let $u$ be the corresponding spin rotation matrix. Then the 
condition for rotational invariance of the hopping process is  
\begin{equation}   \label{uHu-1}
    H^T_{mss'''} = u_{ss'} H^T_{ns's''} u^{-1}_{s''s'''}
\end{equation}
For the on-site hopping amplitudes ($n=m=0$) this implies
\begin{equation}    \label{H0AB}
      H^T_{0ss'} = \epsilon \, \delta_{ss'} 
\end{equation}
For 90$^{\circ}$ rotations about the $n$ axis,  
$u_{ss'} = (\delta_{ss'} + \sigma^n_{ss'})/\sqrt{2}$
where $\sigma^n_{ss'}$ denotes a Pauli matrix. 
Thus, applying (\ref{uHu-1}) to $H^T_{nss'}$ we have $n=m$ and hence 
\begin{equation}    \label{HkAB}
    H^T_{nss'} = \eta \, \delta_{ss'} + \kappa \, \sigma^n_{ss'}   \qquad n=1,2,3
\end{equation} 
Cubic symmetry implies that $\eta$ and $\kappa$ do not depend on $n$.
A further 180$^{\circ}$ rotation ($u=\sigma$) about an axis perpendicular 
to $n$ gives 
\begin{equation}    \label{H-kAB}
    H^T_{(-n)AB} = \eta \, \delta_{ss'} - \kappa \, \sigma^n_{ss'}  \qquad n=1,2,3
\end{equation}
Thus, starting from (\ref{PsiDiff}) we arrive at 
\begin{eqnarray*}
     \frac{\rm d}{{\rm d}t} \psi_s(\vec{x},t) = 
   \epsilon \psi_s(\vec{x},t)  
 &+&  \eta \sum_{n=1}^3 \left(
  \psi_s(\vec{x}+a\hat{n},t) + \psi_s(\vec{x}-a\hat{n},t) \right) \\
 &-&  \kappa \sum_{n=1}^3 \sigma^n_{ss'} \left(
  \psi_{s'}(\vec{x}+a\hat{n},t) - \psi_{s'}(\vec{x}-a\hat{n},t) \right)  
\end{eqnarray*}
We can take the continuum limit $a\to 0$ by expanding the wavefunction,
\begin{equation}    \label{Taylor}
   \psi_s(\vec{x}\pm a\hat{n},t) = \psi_s(\vec{x},t) 
  \pm a \partial_n \psi_s(\vec{x},t) + \frac{a^2}2 \partial_n^2 \psi_s(\vec{x},t) 
  + {\cal O}(a^3)
\end{equation} 
Thus we obtain 
\begin{equation}     \label{preWeyl}
   \partial_t \psi_s(\vec{x},t) = (\epsilon+6\eta)\psi_s(\vec{x},t)
  - 2 a \kappa\,\sigma^n_{ss'}  \partial_n \psi_{s'}(\vec{x},t)
  + \eta a^2 \Delta \psi_s(\vec{x},t) + {\cal O}(a^3)
\end{equation}
We now make another simple but crucial assumption---that all {\it a priori} hopping 
amplitudes are of the same order of magnitude, or at least not dramatically 
different in their order of magnitude. This means, for example, 
that hopping with a spin-flip is about as likely as without it. An argument in favour 
of this assumption is that rotational invariance on a lattice is broken down to cubic 
symmetries; thus orbital angular momentum is certainly not conserved at the length 
scale of the lattice spacing. Spin is also related to the geometrical elements of a 
lattice. In the Dirac-K\"ahler formalism \cite{BennTuck}, for example, spinors arise 
as superpositions of 0-forms, 1-forms, etc.; in the lattice version of the formalism  
\cite{Becher} they are superpositions of sites, links, plaquettes, etc. 

In equation (\ref{preWeyl}) our assumption means that $\kappa$ and $\eta$ are of the 
same order in $a$. Hence, $\eta a^2$ is negligible in comparison to $\kappa a$. 
The terms of ${\cal O}(a^3)$ are irrelevant, too. The general solution of
(\ref{preWeyl}) can then be written in terms of Fourier components of 
$\psi_s(\vec{x},0)$ as
$$
   \psi_s(\vec{x},t) = e^{(\epsilon+6\eta)t} 
   \int {\rm d}^3p \, e^{i\vec{p}\cdot\vec{x}} \, 
          \left[ \cos(2a\kappa |\vec{p}| t) \, \delta_{ss'}  
       - i \sin(2a\kappa |\vec{p}| t) \, \hat{p}\cdot\vec{\sigma}_{ss'} \right] \, 
     \tilde{\psi}_{s'}(\vec{p})
$$
Now consider the case that the Fourier amplitudes $\tilde{\psi}_s(\vec{p})$ decay 
more slowly than exponentially with $|\vec{p}|\to\infty$. If $\kappa$ had an
imaginary part, there would be time evolution factors growing exponentially as
$\exp(2a|\Im\kappa||\vec{p}|t)$ for some of the spin components. This cannot 
be compensated by $\tilde{\psi}_s(\vec{p})$ by assumption, so the Fourier 
integral would diverge immediately for $t>0$ or $t<0$. Hence, $\kappa$ cannot 
parameterize a universal law of time evolution for all spinor wave functions 
unless it is a real number. 
As for the global evolution factor $e^{(\epsilon+6\eta)t}$, an argument about
superpositions of $n$-particle states analogous to the one given in the spin 0
case shows that $\epsilon+6\eta$ must be purely imaginary. Hence, that 
evolution factor can be absorbed by redefining the phase of the spinor wave
function. If we finally identify the speed of light as
\begin{equation}      \label{c}
                       c =  2 a \kappa 
\end{equation} 
we obtain the free Weyl equation    
\begin{equation}                    \label{freeWeyl}
    \partial_t\psi_s(\vec{x},t) = 
    - \, c \, \sigma^n_{ss'}  \partial_n \psi_{s'}(\vec{x},t)
\end{equation} 
One of the reasons for requiring Hamiltonians to be hermitian is to ensure that 
probabilities are conserved. For the systems considered so far, hermiticity was 
redundant because of the symmetries present. Let us finally consider a spin 1/2 
particle in a situation where the symmetries are broken, but only weakly 
in next-to-leading order in $a$. Thus, instead of (\ref{H0AB}) and (\ref{HkAB}),
let the hopping amplitudes be   
\begin{eqnarray}
    H^T_{0ss''} & = & \epsilon \, \delta_{ss'} 
         \left(\delta_{s's''} - a \gamma_{0s's''}(\vec{x},t)\right) 
                                                               \label{hopGen0} \\
    H^T_{nss''} & = & \eta \, \delta_{ss'} + \kappa \, \sigma^n_{ss'} 
        \left(\delta_{s's''} - a \gamma_{ns's''}(\vec{x},t)\right)
                                                               \label{hopGenn}
\end{eqnarray}
with $\epsilon$, $\eta$, $\kappa$ of order $a^{-1}$ as before and $\gamma$,
the spin connection coefficients \cite{BadeJehle,Datta,HehlHeyde}, of order 1.
If no further conditions are imposed on the $\gamma$s we obtain the Weyl equation 
in a space-time with torsion and non-metricity (metric-affine space-time) as currently 
investigated in the context of quantum gravity \cite{HehlNeeman}. Probabilities are 
still {\em covariantly} conserved if we define the probability current as usual 
\cite{BadeJehle,Datta,HehlHeyde} by
$$
     j^{\mu}(\vec{x},t) = \psi_{s}^*(\vec{x},t) ~ \sigma^{\mu}_{ss'}
                          \psi_{s'} (\vec{x},t)
$$
and if we define the tensor connection coefficients $\Gamma^{\mu}_{\nu\alpha}$ 
by requiring the Pauli matrices, which are constant here, to be covariantly constant 
as well \cite{BadeJehle,Datta,HehlHeyde}. In terms of spin-matrix products this 
condition is 
$$
    \Gamma^{\mu}_{\nu\alpha} \, \sigma^{\nu} 
  + \sigma^{\mu} \gamma_{\alpha} + \gamma_{\alpha}^{\dag} \sigma^{\mu} = 0  
$$ 
It then follows by insertion of (\ref{hopGen0}) and (\ref{hopGenn}) in equation 
(\ref{PsiDiff}) and its complex conjugate that
$$
    \partial_{\alpha} j^{\alpha} + \Gamma^{\alpha}_{\beta\alpha} j^{\beta} = 0
$$ 
The tensor metric
$
    g^{\mu\nu} = 
     {\rm tr} \, \left( \sigma^{\mu} \epsilon \sigma^{\nu} \epsilon \right) 
$
is not covariantly constant in this general case. Only by imposing the constraint
$$
     \Re \, {\rm tr} \, (\gamma_{\alpha} \epsilon ) = 0    \qquad \alpha=0,1,2,3    
$$
with $\epsilon=i\sigma_2$ the spin metric would we reproduce the Weyl equation  
in a metric space-time with torsion, expressed in an orthonormal frame \cite{HehlHeyde}.
However, in view of potential applications to quantum gravity, the more interesting 
finding appears to be that the Weyl equation in a metric-affine space-time can be 
obtained on the basis of quantum-mechanical linearity alone, independently of scalar 
products for state vectors.

\end{document}